\documentclass[12pt,preprint]{aastex}
\newcommand{\msun}{$M_{\odot}$}
\newcommand{\ms}{main-sequence star}
\newcommand{\wds}{white dwarf}
\newcommand{\nsix}{NGC 6397}

\newcommand{\sig}{$\sigma$}

\newcommand{\kms}{km\,s$^{-1}$}

\newcommand{\idest}{i.e.~}

\newcommand{\hst}{{\em HST}}
\newcommand{\young}{young}
\newcommand{\old}{old}

\begin{document}

\title{On the Radial Distribution of White Dwarfs in the Globular Cluster NGC 6397\altaffilmark{1}}
\author{
D.S.   Davis\altaffilmark{2},
H.B.   Richer\altaffilmark{2},
I.R.   King\altaffilmark{3},
J.     Anderson\altaffilmark{4},
J.     Coffey\altaffilmark{2},
G.G.   Fahlman\altaffilmark{5},
J.     Hurley\altaffilmark{6},
J.S.   Kalirai\altaffilmark{7,8}
}
\altaffiltext{1}{Based on observations with the NASA/ESA Hubble Space Telescope, obtained at the Space Telescope Science Institute, which is operated by the Association of Universities for Research in Astronomy, Inc., under NASA contract NAS5-26555. These observations are associated with proposals GO-10424 and GO-8679.}
\altaffiltext{2}{Department of Physics \& Astronomy, University of British Columbia, 6224 Agricultural Rd., Vancouver, BC, V6T 3B4, Canada; sdavis@astro.ubc.ca}
\altaffiltext{3}{Department of Astronomy, University of Washington, Seattle, WA, USA}
\altaffiltext{4}{Department of Physics \& Astronomy, Rice University, Huston, Texas, USA}
\altaffiltext{5}{Herzberg Institute of Astrophysics, National Research Council of Canada, Saanich, BC, Canada}
\altaffiltext{6}{Centre for Astrophysics \& Supercomputing, Swinburne University of Technology, Hawthorn, Australia}
\altaffiltext{7}{University of California at Santa Cruz, Santa Cruz, CA, USA}
\altaffiltext{8}{Hubble Fellow}

\begin{abstract}
 We have examined the radial distribution of white dwarfs over a single HST/ACS field in the nearby globular cluster NGC 6397. In relaxed populations, such as in a globular cluster, stellar velocity dispersion, and hence radial distribution, is directly dependent on stellar masses. The progenitors of very young cluster white dwarfs had a mass of $\sim 0.8$ \msun, while the white dwarfs themselves have a mass of $\sim0.5$ \msun. We thus expect young white dwarfs to have a concentrated radial distribution (like that of their progenitors) that becomes more extended over several relaxation times to mimic that of $\sim 0.5$ \msun\ main-sequence stars. However, we observe young white dwarfs to have a significantly extended radial distribution compared to both the most massive main sequence stars in the cluster and also to old white dwarfs. 
\end{abstract}
 

\section{Introduction}
The large proper motions of some pulsars indicate that they have space velocities of hundreds of kilometers per second, which were presumably imparted to them when they became neutron stars \citep[for recent reviews see][]{ppr05,rom05}. The question naturally arises, could something of this sort have happened to \wds s? They are, of course, not observed to have comparably high velocities, and their births do not involve anything as energetic as a supernova, but there is already a hint in the literature that white dwarfs may have higher velocities than their progenitors did. \cite{wei97} and \cite{wil02} have observed that the numbers of white dwarfs observed in open clusters are lower than expected---a natural explanation being that a fraction of the white dwarfs have escaped from the cluster. To investigate this question further, \cite{flb03} performed N-body simulations and found that a kick of magnitude approximately twice that of the cluster velocity dispersion (several \kms)  would deplete a loosely bound open cluster of almost all its \wds s.

The hypothesis that white dwarfs begin their lives with a small velocity excess leads to scenarios that are particularly conducive, for dynamical reasons, to testing in a globular cluster. The progenitors of the newest \wds s in a typical globular cluster had a mass of $\sim0.8$ \msun, but they have become \wds s with a mass of $\sim0.5$ \msun. Assuming a quiescent birth, they begin their lives with the spatial distribution that is appropriate for their progenitors, but over the course of a relaxation time---of the order of a few times $10^8$ years---they will acquire the more extended spatial distribution that goes with their new, lower mass. However, if \wds s are given natal kicks, they then start their lives with a velocity dispersion that is too large for their original spatial distribution, and in the course of a crossing time---only a million years or so---they acquire a more extended spatial distribution, perhaps even more extended than they will have after relaxation has made their spatial distribution correspond to their new masses. In a typical globular cluster, stars well down the white dwarf cooling sequence have been \wds s for dozens of relaxation times, while the youngest \wds s have entered that stage very recently. Moreover, these clusters have hundreds of \wds s, so that the test can be made in a statistically significant way. Finally, because globular clusters are old stellar populations, the mass difference between the progenitors of the youngest and oldest \wds s in our sample is $<0.09$ \msun. Though the \wds\ initial-final mass relation is uncertain \citep{fwl05,krr05}, the mass difference between the \wds s must certainly be much less than the mass difference between their progenitors. This allows us to treat all \wds s as essentially of equal mass. 

The question then is, do young \wds s in a globular cluster have a spatial distribution that is more extended than that of their progenitors? Until recently, the problem with addressing this question has been that \wds s in a globular cluster, particularly the old, fainter ones,  are at a magnitude where they cannot be easily observed. Today, however, deep imaging with \hst\ is able to reach faint enough to image the entire white dwarf sequence in very nearby globular clusters \citep{rab06,hab07}. We will show here that in NGC 6397 the radial distribution of its young \wds s is significantly more extended than that of their progenitors and also old \wds s. 
\cite{dav06} presented preliminary results on the white dwarf distribution in this cluster and   \citep{hey07a,hey07b} then used Monte Carlo simulations to generate radial distributions of stars with kicks at birth and found that a kick of the order of 1.8 times the velocity dispersion of the cluster giants was required to account for the effect seen by \cite{dav06}.

\section{Observations}
In \nsix\ we used {\it HST's} ACS Wide Field Channel to image a single field centered $5\arcmin$ SE of the cluster center for 126 orbits \cite[GO-10424, for details see][]{rab06}. The exposure time was divided between the F814W and F606W filters. The field ranges from $179\arcsec$ to $391\arcsec$ from the cluster center, corresponding to $1.3$--$2.8$ half-mass radii \citep{har96}. Proper motions, for the elimination of field stars, were measured on archival WFPC2 images. We were able to extend our magnitude limit considerably by blind measurement of the WFPC2 images at every point where the ACS images told us that there was a star, but the shorter exposure time of the WFPC2 images was still what set the limit of our completeness.  Even so, artificial-star tests showed that at the magnitude that corresponds to a white dwarf age of 5 Gyr our completeness was still 92\%.
Table \ref{prop.tab} summarizes the major properties of NGC 6397.

\begin{figure}
\plotone{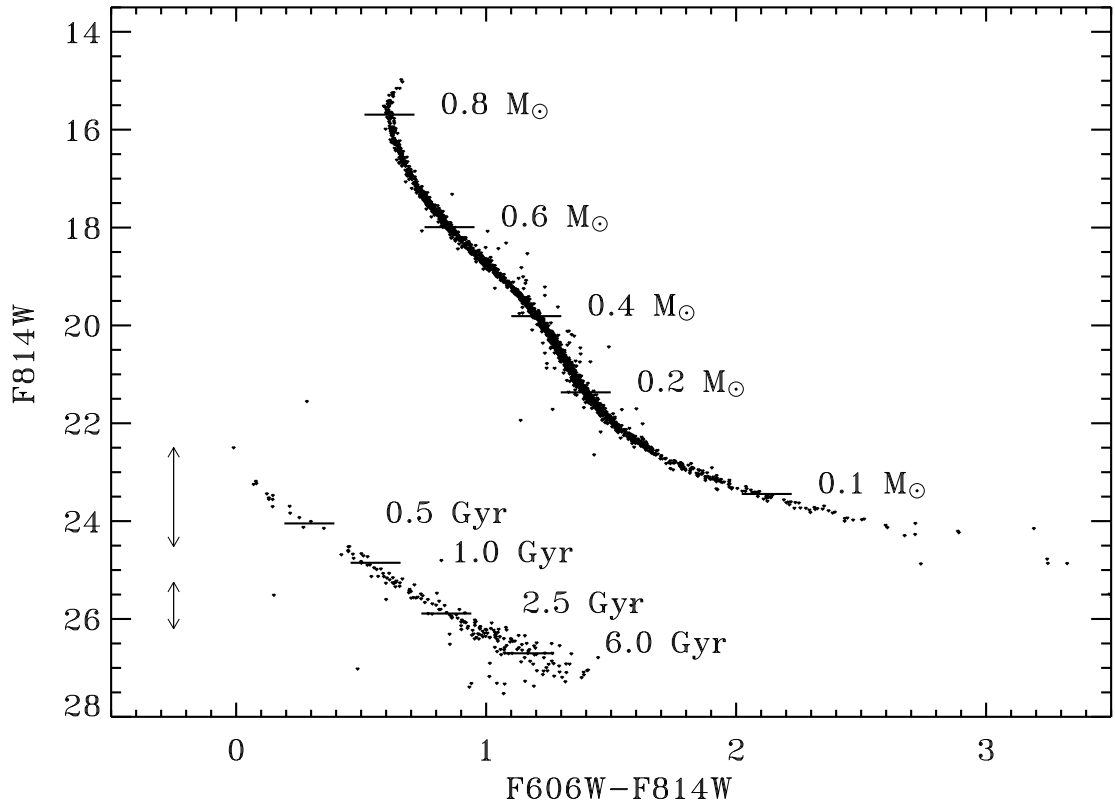}
\caption{The CMD of NGC 6397. The \wds\ ages, according to the Wood/Bergeron white dwarf cooling models, and the main-sequence masses, as determined from models of Baraffe, are indicated. The magnitude range of the old (lower) and young (upper) \wds s are shown with arrows.\label{cmd.fig}}
\end{figure}

The essence of this {\em Letter} is to compare the radial distributions of \wds s of various ages with those of \ms s of various masses. Neither mass nor age is directly observable; in what follows we will need models of both \ms s and cooling \wds s. To assign a mass to each \ms, we use main-sequence models of \cite{bca97}, with the metallicity appropriate for NGC 6397. To assign an age to each \wds, we use a set of cooling models that were developed using interiors from \cite{woo95}  and atmospheres from \cite{bwb95}.  Our cooling models are for $0.5$ \msun\ stars with pure carbon cores and thick hydrogen-rich atmospheres. Shown in Figure \ref{cmd.fig} is the colour-magnitude diagram (CMD) of NGC 6397 with the main-sequence masses and white dwarf cooling ages superimposed.

\section{Radial distributions}
To test whether two distributions differ significantly, we make use of two statistical tests:\ the familiar Kolmogorov-Smirnov (KS) test, which uses the maximum difference between the normalized cumulative distributions, and the Wilcoxon rank-sum (RS) test (equivalent to the Mann-Whitney U-test), which compares rankings within the merged, sorted distribution. Each of the tests gives the probability that two random samplings from the same parent population will show differences as large as the observed ones, though the RS test (because it can be applied as a one-tailed test) is generally more sensitive.

\begin{figure}
\plotone{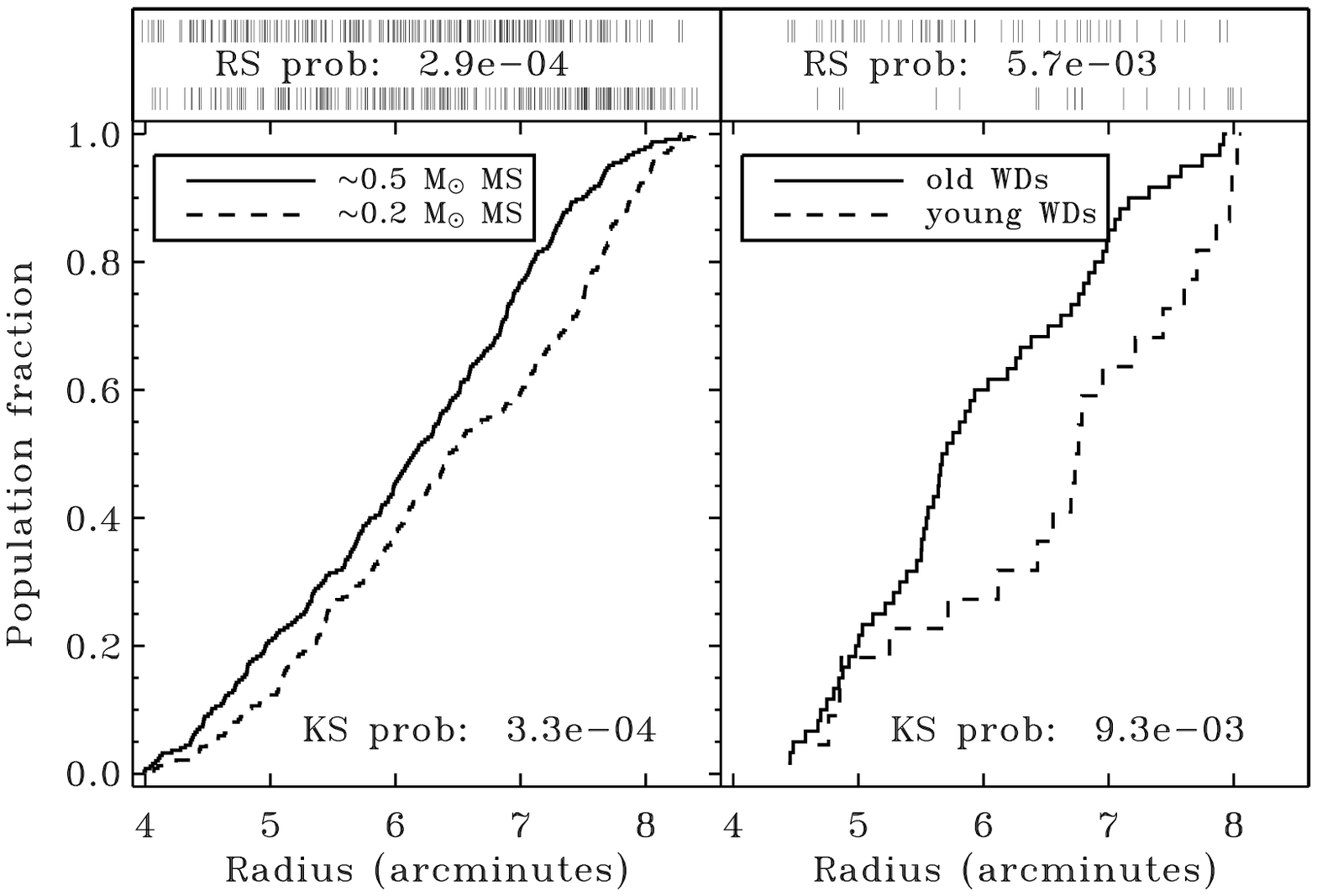}
\caption{Left section: A comparison of the radial distributions of $0.5$--$0.6$ \msun\ and $0.15$--$0.20$ \msun\ \ms s for \nsix. The upper panel shows the RS test and the lower panel the KS test. The vertical lines in the upper panel indicate the radial position of each star ($0.5$--$0.6$ \msun\ above and $0.15$--$0.20$ \msun\ below). Right section:  A comparison of the radial distributions of the \young\ and \old\ \wds s. The panels are the same as in the left figure with the vertical bars denoting the locations of the old (upper) and young (lower) \wds s. Surprisingly, the \young\ \wds\ population is {\em less} centrally concentrated than the \old\ population. \label{cum.fig}}
\end{figure}

Mass segregation is known to be present in  \nsix\ \citep{ksc95,atm04}. We test the sensitivity of our data set to mass segregation by comparing the radial distributions of $0.15$--$0.2$ \msun\ \ms s with $0.5$--$0.6$ \msun\ \ms s (\idest\ the presumed true mass of the \wds s). Figure \ref{cum.fig}, left section,  shows the cumulative radial distributions of these two groups, and the power with which the KS and RS tests can distinguish between them. Our data are clearly able to detect mass segregation, even over the limited radial extent of our field.

The white dwarfs that have had time to undergo dynamical relaxation should have a distribution similar to that of the $\sim0.5$ \msun\ \ms s, while the unrelaxed ones should have a more centrally concentrated distribution, like their more massive progenitors. However, NGC 6397 exists in the Galactic potential, and periodic changes in the lumpy potential could affect the relaxation process. In fact, the orbit of \nsix\ has a disk-crossing time of $\sim 100$ Myr \citep{mvb06,kar07}, and one might expect disk shocking to play a role in the internal cluster dynamics. However, in relaxed clusters, such as NGC 6397, we expect these effects to be most important close to the tidal radius, and insignificant at the half-mass radius \citep{glo99}.

To study the distributions of the \wds s, we need a ``young'' and an ``old'' sample. (When we refer to age here, we mean the time since the star became a \wds.) We want all the stars of the \young\ sample to be young enough not to have experienced appreciable relaxation since they acquired the lower mass that goes with being a \wds. Therefore, we need to look at the relaxation time in our region---bearing in mind, however, that relaxation time is only an approximate time scale rather than an exact quantity. The \cite{har96} catalog gives a relaxation time of $0.29$ Gyr at a half-mass radius of  $2\farcm3$ for \nsix. Given that even the inner edge of our field is at a radius where the star density is lower than it is at  $2\farcm3$, our stars should remain unrelaxed for $\sim$ 1 Gyr.  

For choice of the \old-sample boundaries there are two considerations. The lower boundary in age should be high enough so that the stars have had enough time to have relaxed to the distribution that is appropriate for their white dwarf mass. The upper age limit must be low enough to ensure that the completeness at the magnitude of the faintest \wds\ is not a function of crowding, and therefore radial position. We take the \old\ group to be from $1.4$--$3.5$ Gyr (62 stars). 

For the choice of the \young-sample boundaries there is a different set of considerations. The lower boundary should not be less than a crossing time. However, this timescale is of the order of Myrs---shorter than the youngest inferred \wds\ age in our sample---so this boundary can be set to zero Gyr. In principle it would be advantageous to set the upper boundary to select only the very youngest (\idest\ age {\em much} less than a relaxation time) \wds s for this sample, thereby ensuring that the young \wds s would have had very little time to undergo interactions with other stars.  In practice, it is necessary to set the upper boundary to an age comparable to a relaxation time in order to increase the sample size, while maintaining a sample largely free from the effects of relaxation. We choose a \young\ sample that goes from $0.0$--$0.8$ Gyr (22 stars). We note that we are in no danger of inducing a false signal by choosing an upper boundary that includes some relaxed \wds s, but we will rather dilute any signal that is truly there.  

While our definitions of \young\ and \old\ were guided by physical principles and observational constraints, they remained somewhat arbitrary.  In fact, there is some sensitivity to the upper boundary of the young group.  If that boundary is placed anywhere from 0.4 to 0.8 Gyr the result remains strongly significant, increasing the boundary beyond this decreases the significance. Figure \ref{cum.fig}, right section, compares the radial distributions of the \young\ sample with the \old\ one in NGC 6397.

Contrary to what one would expect supposing a quiescent birth scenario, the younger white dwarfs tend to be farther from the cluster center than old white dwarfs. This  suggests that they have a higher velocity dispersion than old \wds s. However, this is {\it not} the key point. The most direct test would be to compare the radial distribution of young \wds s with that of their progenitors. Obviously, this test is impossible, as \ms s of the mass of the progenitors no longer exist in the cluster. Instead we will test the \wds\ distributions against a series of main-sequence mass samples. Assuming that all \ms s are relaxed, and that the trend is monotonic with mass, if we can show that the radial distribution of the young \wds s is inconsistent with the radial distribution of the highest mass \ms s that we observe, we have shown that young \wds s are not drawn from the same population as their progenitors.  
 
\begin{figure}
\plotone{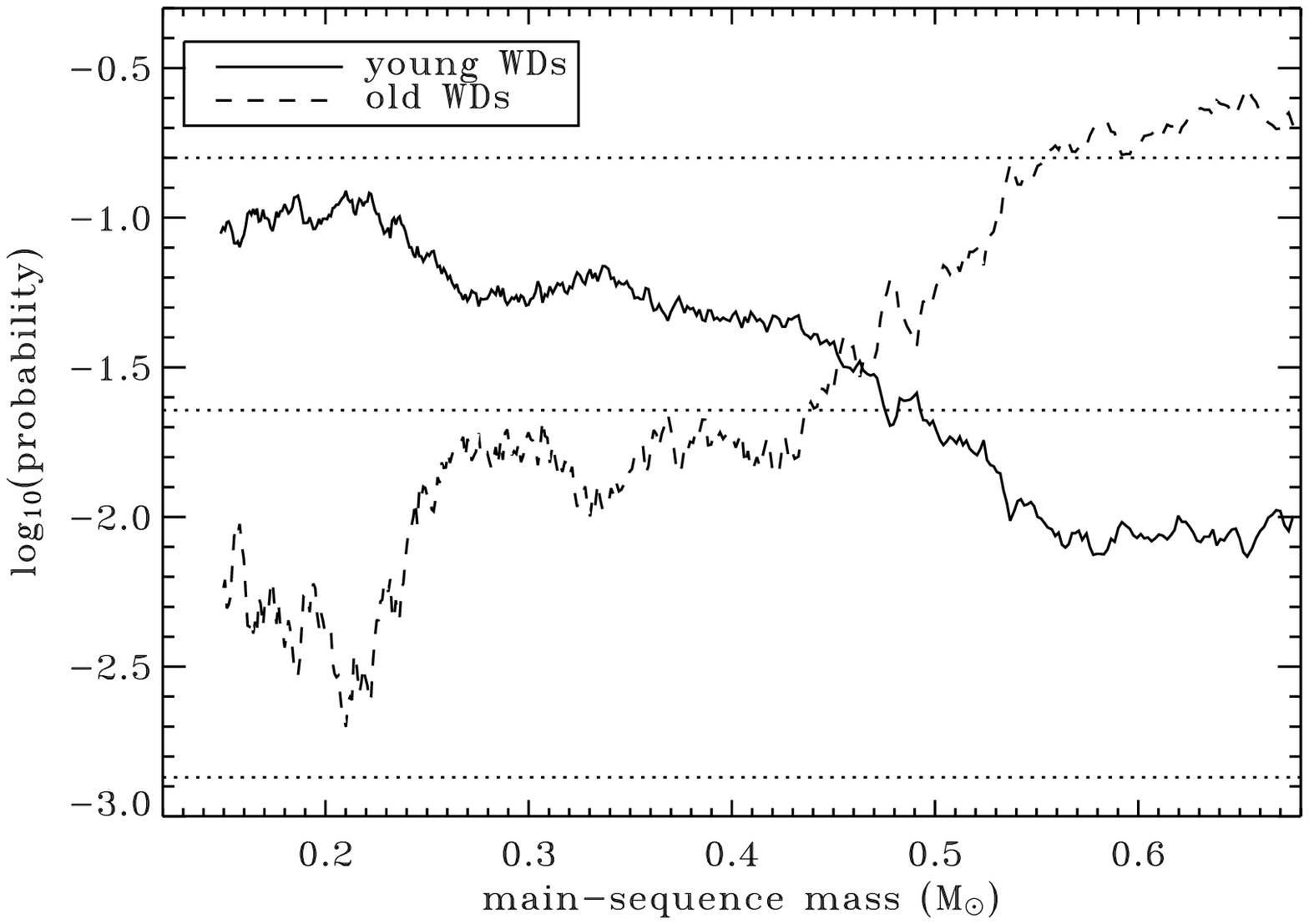}
\caption{The RS probability that the radial distribution of a given \wds\ population could be drawn from the same parent distribution as the radial distribution of a given mass of \ms\ for NGC 6397. The dashed lines indicate the equivalent of one-tailed 1\sig, 2\sig, and 3\sig\ results. Note: the mass value of the points are the mean values of the mass bins. Adjacent points making up the lines are not independent, and are composed of approximately 90\% the same stars.
\label{zzz.fig}}
\end{figure}

We use the RS test to calculate the probability that the radial distributions of the \young\ and \old\ \wds s could come from samplings of the same parent distribution as \ms s of various masses; the results are shown in Figure \ref{zzz.fig}. Setting the acceptance level at 0.05 (2\sig\ in a one-tailed normal distribution), we can {\em reject} the hypothesis that the young \wds s have a radial distribution consistent with that of \ms s more massive than $\sim0.5$ \msun\ in NGC 6397. Furthermore, we note that we cannot reject the hypotheses that these same stars have radial distributions consistent with that of lower mass \ms s. The rejection of the former hypothesis leads us to conclude that the young \wds s in NGC 6397 have a radial distribution more extended than that of their progenitors. Lending confidence to our method, Figure \ref{zzz.fig} also shows that the radial distribution of the old \wds s is not inconsistent with that of \ms s of their supposed mass, but is inconsistent with that of low mass ($M<0.25$ \msun) \ms s.

\section{Are white dwarfs born with a kick?}
We have shown that the radial distribution of young \wds s in NGC 6397 is more extended than would be expected given a quiescent birth. Other than a possible kick at birth, is there an alternate explanation for this observation?  One might imagine a mechanism related to the white dwarfs being in binary
systems, either through binary disruption or the inhibition of white dwarf formation through common envelope evolution or mergers. However, the binary fraction in \nsix\ is low \citep{cb02}, and any explanation invoking binaries is implausible. Alternatively, the newly born \wds s could undergo very close encounters during their first passage through the inner regions of the cluster with their reduced masses, and have their velocities increased to greater-than-equilibrium values through stellar interactions alone. We have produced N-body simulations that accurately model both binary evolution and stellar encounters, but omit natal kicks; we do not observe the young \wds s to have extended radial distributions in these simulations.  A full description of these N-body models can be found in \cite{hur07}. \nsix\ has a collapsed core, and one may consider if interactions with the central density cusp could somehow create this effect. However, the white dwarf progenitors experienced the same potential as white dwarfs, so it is difficult to see how this could have an effect.

Assuming the existence of a natal kick, we can estimate its size in the following way. At the radial distances that we are considering, the measured velocity dispersion of $\sim 0.8$ \msun\ giants in \nsix\ is $3.3$ \kms\ \citep{pm93}. In accordance with equipartition of energy in a relaxed system, we expect the velocity dispersion of stars of mass $M$ to be given by \sig$_{\rm vel}=\sigma_{\rm giant}\sqrt{M_{\rm giant}/M}$. Thus, \sig$_{\rm vel}$ for $0.2$ \msun\ stars should be about twice that of the giants. Because velocity dispersions add in quadrature, the magnitudes of impulsive kicks that would transform the radial distribution of a $0.8$ \msun\ population to that of a $0.5$ \msun\ or $0.2$ \msun\ population are 3 \kms\ and 6 \kms, respectively. These values are comparable to the velocity dispersion of a typical globular cluster, but still well below their escape velocities. 

The fact that we are examining a relatively small range of radii, and that we have a small number of stars in the \wds\ sample, prevents us from putting tight constraints on the velocity dispersion of the young \wds s. However, we note that the young \wds s have an extended radial distribution, implying an extra source of velocity dispersion. The most natural explanation for this observation is the existence of a natal kick. We leave it to the theorists to search for a detailed mechanism, but note that asymmetric mass loss while the stars are on the asymptotic giant branch stages could generate a kick of the inferred magnitude \citep{flb03}. 
 
DSD would like to thank J.~Wall for useful conversations and the UBC-UGF for funding. HBR is generally funded by  NSERC, but support for this project was also provided through UBC by the Vice President for Research, the Dean of Science, the NSERC Emergency Fund, and the Department of Physics and Astronomy. HBR also thanks the Canada-US Fulbright Committee for the award of a Fulbright fellowship. JA, IRK, and JSK  received support from NASA/HST through grant GO-10424. JSK is supported by NASA through Hubble Fellowship grant HF-01185.01-A, awarded by the Space Telescope Science Institute, which is operated by the Association of Universities for Research in Astronomy, Incorporated, under NASA contract NAS5-26555.

\clearpage

 \begin{deluxetable}{lc}
\tablecolumns{2}
\tablewidth{0pc}
\tablecaption{NGC 6397: Major Properties \label{prop.tab}}
\tablehead{
\colhead{Property}  & \colhead{Value}
 }
 \startdata
 
 Distance Modulus (true)   & $12.03\pm0.06^1$   \\
A$_{\rm F814W}$           &  0.33$^{1,2}$      \\
$[$Fe/H$]$                & $-2.02\pm0.07^3$   \\
R$_{\rm half\ mass}$      & 140$\arcsec^4$      \\
t$_{\rm half\ mass}$      & 0.29 Gyr$^4$        \\
Observed field radius     & 179--391$\arcsec$   \\
\enddata
\tablerefs{
(1)  \cite{hab07};
(2)  \cite{sjb05};
(3)  \cite{ki03};
(4)  \cite{har96}
}
 \end{deluxetable}

\clearpage

\clearpage

\clearpage

\end{document}